\documentclass[twocolumn,showpacs,preprintnumbers,amsmath,amssymb,prb]{revtex4}
\usepackage{graphicx}
\usepackage{dcolumn}
\usepackage{bm}
\begin{document}
\title{Failure of $t-J$ models in describing doping evolution of
spectral weight in x-ray scattering, optical and
photoemission spectra of the cuprates}
\author{R. S. Markiewicz, Tanmoy Das, and A. Bansil}
\address{Physics Department, Northeastern University, Boston MA
02115, USA}
\date{\today}
\begin{abstract}
We have analyzed experimental evidence for an anomalous transfer of
spectral weight from high to low energy scales in both electron and hole
doped cuprates as a function of doping. X-ray scattering, optical and
photoemission spectra are all found to show that the high energy spectral
weight decreases with increasing doping at a rate much faster than
predictions of the large $U-$limit calculations. The observed doping
evolution is however well-described by an intermediate coupling scenario
where the effective Hubbard $U$ is comparable to the bandwidth.
The experimental spectra across various spectroscopies are
inconsistent with fixed-$U$ exact diagonalization or quantum Monte Carlo calculations, and
indicate a significant doping dependence of the effective $U$ in the
cuprates.
\end{abstract}
 \maketitle\narrowtext

\section{Introduction}
The key to unraveling the mechanism of cuprate superconductivity is to
ascertain the effective strength of correlations since pairing is widely
believed to arise from electron-electron interactions rather than from the
traditional electron-phonon coupling. Two sharply different scenarios have
been proposed and remain subject of considerable debate. One viewpoint
holds that $U\gg W$ where $U$ is the Hubbard $U$, and $W\sim 8t$ is the
bandwidth with hopping parameter $t$. In this case, a `pairing glue' is
not necessary as the pairs are bound by a superexchange interaction
$J=4t^2/U$, and the dynamics of the pairs involves virtual excitations
above the Mott gap set by the energy scale $U$.\cite{anderson} In the
opposing view, $U\sim W$, and pairing is mediated by a bosonic `glue',
which originates from antiferromagnetic (AFM) spin
fluctuations\cite{maierDSC,scalapino}. It is clear thus that the
determination of the size of the effective $U$ and its variation with
doping are essential ingredients for understanding the mechanism of
superconductivity as well as the magnetic phase diagram of the cuprates.

Since the electronic dispersion at half filling has a gap of magnitude
$\sim U$ which is clearly visible in x-ray absorption (XAS), angle-resolved photoemission (ARPES), and optical
spectra, one way to estimate the size of $U$ is to follow the evolution of
high-energy spectral weight as a function of doping.  Whereas for a
conventional band insulator the spectral weights of the bands above and
below the gap are independent of doping, this is not the case for a Mott
insulator.  In the latter case, for $U\rightarrow\infty$, removing one
electron creates {\it two} low energy holes -- one from the lower Hubbard
band (LHB), but a second one from the upper Hubbard band (UHB), since
without an electron on the atom there is no $U$-penalty in adding an
electron.  Paradoxically, as $U$ decreases, the rate of this anomalous
spectral weight transfer (ASWT) actually increases\cite{EMS}.  For
infinite $U$ double occupancy (DO) is always forbidden, so no matter how
few electrons are in the LHB, there will be an equivalent number of holes
in the UHB.  In contrast, for smaller $U$ values DO is reduced
collectively, via long-range magnetic order.  As the magnetic order
disappears at a quantum critical point\cite{tanmoy07,tanmoy08}, a much higher degree of DO is
restored, and the UHB can completely vanish.

Hence, by measuring the high energy spectral weight (HESW) as a function
of doping, we can estimate the degree of correlation in cuprates. Here we
quantify these results for XAS, ARPES, and optical measurements, and demonstrate that the doping
evolution of ASWT is similar across all these spectroscopies for both
electron and hole doped cuprates. Moreover, the observed doping evolution
is inconsistent with large $U$ values, and also with fixed-$U$ Hubbard
model calculations, but it is consistent with a doping-dependent effective
$U$ of intermediate strength $U\lesssim W$.

This paper is organized as follows. In Section II we explain how to
quantify the rate of ASWT with doping, and show that similar rates are
found for several different spectroscopies. In Section III we show that
these rates are consistent with an intermediate coupling model of the
cuprates. A discussion of the results is given in Section IV, and
conclusions are presented in Section V.

\section{Quantifying Anomalous Spectral Weight Transfer}
We motivate a definition of the rate of ASWT with doping in Fig.~1(a).  In a strongly
correlated system, removing $x$ electrons produces ($1+x$) states above
the Fermi level, which are distributed between $p\ge 2x$ low energy
(in-gap) states and $W_{UHB}=1+x-p$ states in the UHB.  Then the ASWT can
be quantified by the coefficient $\beta$, defined such that in this
process the weight of the UHB reduces to $W_{UHB}=1-\beta x$ and the low
energy holes gain weight by $p=(1+\beta)x$. The value of $\beta$ is found
theoretically to depend on $U$ such that $\beta =1$ for a very strongly correlated
($U\rightarrow\infty$) Mott insulator,  while reducing $U$ leads to
larger values of $\beta$.  Figure~1(b) illustrates a variety of
calculations of the HESW vs doping. Exact diagonalization (ED)
calculations on small clusters\cite{EMS,foot3} (dashed lines) find $\beta
=1$ for the $t-J$ model or for a $U\rightarrow\infty$ Hubbard model,
$\beta\simeq1.5$ (at low doping) for $U=10t$ and $\beta\simeq2.0$ for
$U=5t$.  Shown also in Fig.~1b are QMC results for $U=8t$, $t'=0$ [where
$t$ and $t'$ are hopping parameters]\cite{jarrell,maier}, which are
consistent with the ED results.  Hence, $\beta =1$ confirms strong
correlations and the `no double occupancy' (NDO) hypothesis, while a
faster falloff ($\beta >1$) indicates otherwise, and supports a real gap
collapse model ($W_{UHB}\sim 0$ at $x=1/\beta$).  For an electron doped
system the HESW is associated with the LHB, and is described by the mirror
image of Fig~1(a) with respect to $E_F$.

\begin{figure}
\rotatebox{270}{\scalebox{0.62}{\includegraphics{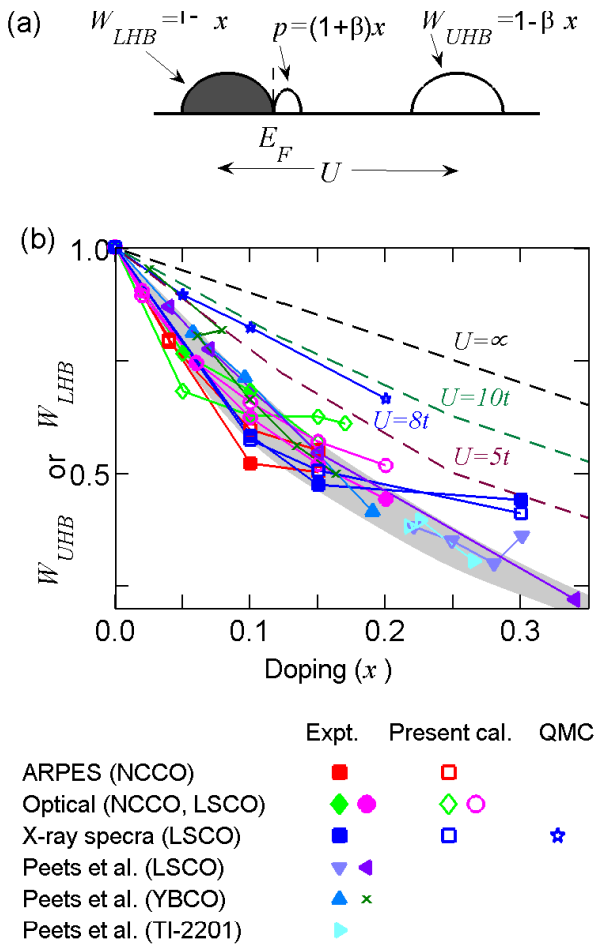}}}
 \caption{(color online) (a)
Schematic diagram of ASWT for hole doped cuprates. (b) Estimates
of $W_{UHB}$ (for hole doping) and $W_{LHB}$ (for electron doping)
from various experimental results (see legend)
\cite{nparm,Pellegrin,onose,onoseprb,uchida,peets} are compared
with our theoretical results (open symbols of same color). Dashed
lines of various colors show exact diagonalization calculations
for different values of $U$ taken from Ref.~\onlinecite{EMS}.  QMC
results\cite{jarrell,maier} from Fig~4 are plotted as blue stars.
All curves are normalized to $W_{UHB}\rightarrow1$ at half
filling. } \label{aswt}
\end{figure}

Shown also in Fig.~1(b) is our key result, the HESW of a variety
of cuprates as a function of doping, extracted from a number of
spectroscopies. The results are
strikingly similar over a variety of spectroscopies, as expected,
but also over several families of cuprates for both electron and
hole doping.  Shown in Fig.~1(b) are XAS results on
La$_{2-x}$Sr$_x$CuO$_4$ (LSCO)\cite{Pellegrin}, ARPES on
Nd$_{2-x}$Ce$_x$CuO$_{4\pm\delta}$ (NCCO)\cite{nparm}, and optical
absorption on both NCCO\cite{onose,onoseprb} and
LSCO\cite{uchida}, compared with additional XAS data for LSCO,
YBCO and Tl-2201 from Ref.~\onlinecite{peets}. All experimental
measures of HESW find a rapid falloff of the spectral weight with
doping, and at low doping decrease almost linearly with doping
with approximately the same slope of $\beta\simeq3.7$, consistent
with $U_{eff}<5t$, suggesting that the cuprates are far from the
strong correlation limit. The observed falloff supports a real gap
collapse at $x_{UHB}\sim 1/\beta =0.27$.  Notably, the value of
$U_{eff}$ is incompatible with the measured gap at half filling.
For example, optical spectra find a gap consistent with $U\sim
8t$, but the HESW calculations for fixed $U=8t$ are far from the
experimental results. On the other hand, the experimental
data can be explained by intermediate coupling model
calculations\cite{markiewater,tanmoysw,susmita} with a doping
dependent effective $U$. The calculated results are plotted in Fig~1(b)
as open symbols of same color as the corresponding experimental data.

Since the HESW is an intrinsic property of the electronic structure of
cuprates, it should show up in all spectral probes, and Fig.~1(b) confirms
this.  However, it is important to realize that the ASWT will play out
quite differently in different spectroscopies.  First, as is clear from
Fig.~1(a), there is a strong electron-hole asymmetry to the effect: the
changes will be much smaller in the Hubbard band which lies at the Fermi
level.  Hence, for maximum sensitivity to ASWT in a hole-doped cuprate,
the probe should be sensitive to empty states, and to filled states for
electron-doped cuprates.  Thus, ARPES\cite{arpes} or X-ray emission
spectroscopies are well-suited for studying ASWT in electron-doped
cuprates, while XAS is appropriate for hole-doped cuprates.
Optical\cite{tanmoysw} and resonant inelastic x-ray
scattering\cite{MBRIXS,rixs} studies would work for both cases, as they
measure a joint density of states. On the other hand, Compton
scattering\cite{compton} and positron annihilation\cite{positron} will not
be sensitive to ASWT because these spectroscopies measure only the total
spectral weight of occupied states, but not how this spectral weight gets
rearranged in energy with doping. In principle, scanning tunneling
microscopy\cite{stm} could follow either sign of charge, but would require
a wide energy range, $\sim$2eV to see the full effect.

\begin{figure*}[htop]
\rotatebox{270}{\scalebox{0.65}{\includegraphics{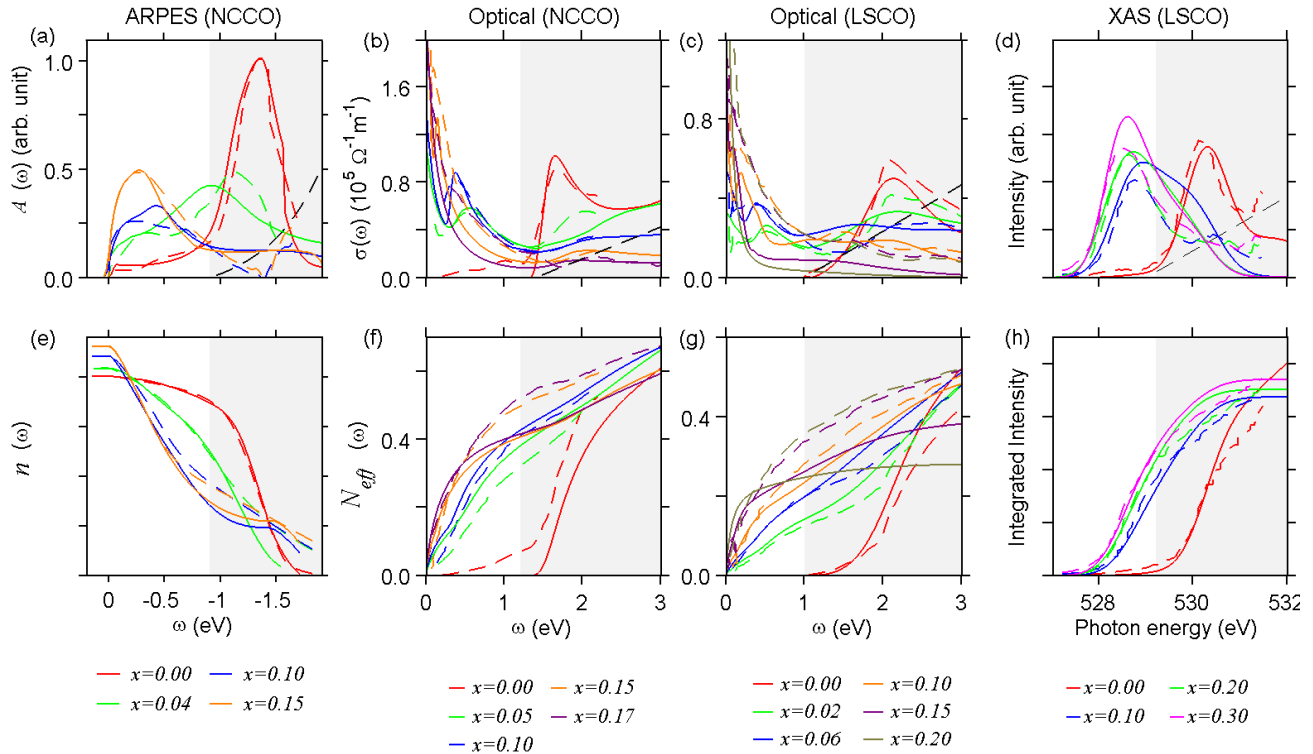}}}
\caption{(color online)
(a) ARPES spectra along the nodal direction of NCCO for various
dopings.\cite{nparm} (b-c) Optical conductivity of NCCO\cite{onose} and
LSCO\cite{uchida}. (d) K-edge XAS results\cite{Pellegrin,peets} are
compared with our theorerical DOS (broadened with experimental resolution
of 0.4eV). All DOS curves are shifted by a doping independent x-ray edge
energy value of 528.4~eV. (e-h) Integrated spectral weights corresponding
to the spectra after subtracting a background associated with
higher-energy bands\cite{spagh}, shown as black dashed lines in frames
(a-c). (e) Integrated ARPES spectral weight (integrated around a small
momentum window to mimic the experiment and averaged over nodal and
antinodal directions), normalized to $(1 + x)$ at $E_F$. (f-g) Effective
number of electrons (Eq.~\ref{eq:2}) calculated from the optical spectra
for NCCO and LSCO. (h): Integrated XAS intensity. In all frames, dashed
lines are experimental data\cite{nparm,onose}; solid lines of same color
are the present calculations, while the edge of the shaded region marks
the crossover energy $\omega_c$, discussed in the text.
}
\label{sumrulef}
\end{figure*}

Figure~2 illustrates how the experimental ASWTs of Fig.~1(b) were
extracted.  The ARPES, optical and XAS data are shown as dashed lines in the upper
panel of Fig.~2, and the corresponding integral, representing the
electron number $n(\omega)$, in the lower panel of Fig.~2.
In each spectrum, the UHB [LHB for electron doping] is denoted by a
gray shaded region, and the $W_{UHB}$ [or $W_{LHB}$], Fig.~1(b),
is defined as the integrated density over that region, starting
from a cutoff frequency $\omega_c$, taken as independent of doping. A complication is
involved in comparing our one-band calculation with the
experimental data, in that the antibonding band in cuprates lies
near to other bands, and the role of the latter must be
disentangled before the spectral weight can be estimated. At high
energies, we subtract off a background from the
experimental spectra associated with interband transitions to
higher-lying bands not included in the present one band
calculations.\cite{spagh,comanac2,tanmoysw,comanac,towfiq}. We use a
doping-independent background contribution shown as black
dashed lines in Figs.~1(a-c).  In all cases we compare the data with
calculations based on the QP-GW model\cite{tanmoysw,tanmoyZ} (solid lines),
discussed below.

For electron doping, ARPES can detect the full LHB and hence determine
$W_{LHB}$ to the extent that matrix element effects are doping
independent.\cite{susmita} The ARPES results for NCCO are compared with
our theoretical results in Fig.~2(a). At half-filling the energy
distribution curve (EDC) along the nodal direction shows the so-called
charge-transfer gap from the Fermi level to the LHB. A
significant redistribution of spectral weight is evident at $x=0.04$ as
the LHB approaches the Fermi level and by $x=0.10$, virtually all of the
spectral weight of the LHB has shifted to the vicinity of the Fermi level.
The top of the LHB crosses $E_F$ at $x\simeq0.15$, forming a hole pocket, and the
spectral weight near $E_F$ undergoes an abrupt increment. To extract the
total spectral weight associated with the LHB we have integrated the
spectral weight from $-1.9$~eV to $\omega$, Fig.~2(e). ARPES data are
available along only two high-symmetry directions, so we take their
average as representative of the net spectral weight, and at each doping
normalize $n(\omega)$ to $(1+x)$ at $E_F$.

In Fig.~2(b,f), analysis of the HESW in NCCO based on the optical
absorption spectra\cite{uchida,onoseprb} proceeds
similarly\cite{tanmoysw}.  There is a large Mott gap below 2~eV in the
undoped material, but with doping there is a strong transfer of spectral
weight from the gap to low energy features -- the Drude peak and the
midinfrared (MIR) peak -- with an isosbetic (equal absorption) point
around $\omega\sim1.3$~eV.  As a measure of the HESW, the effective
electron number (per Cu atom) is obtained as
\begin{equation}\label{eq:2}
N_{eff}(\omega)=\frac{2m_0V}{\pi e^2\hbar}\int_{-\infty}^{\omega}
\sigma(\omega^{\prime})d\omega^{\prime},
\end{equation}
where $m_0$ and $e$ are the free electron mass and charge and $N$ is the
number of Cu-atoms in a cell of volume $V$. The weight of the LHB is
extracted as $W_{LHB}=1+x-N_{eff}(\omega_c)$. A similar analysis of
$W_{UHB}$ was carried out on LSCO spectra\cite{uchida} in Fig.~2(c,g),
and the results included in Fig.~1(b). These optical results are
consistent with the analysis of Comanac, {\it et al.}\cite{comanac}. It is
interesting to note that the $\omega_c$ which separates the high-energy
Hubbard bands and the low-energy in-gap states coincides with the
isosbetic or equal absorption point in the optical spectra i.e., the
residual charge-transfer gap.

For hole doping, $W_{UHB}$ was determined by x-ray absorption spectroscopy
(XAS)\cite{Pellegrin,peets}, which detects the empty states above the $E_F$.  In
this spirit, we compare the measured XAS spectra with the calculated
empty-state DOS in Figs.~2(d) and 2(h). The behavior of the spectral
weight transfer is very similar to the ARPES result for NCCO in Fig.~2(a).

The overall similarity of the doping dependence of the excess electron [or hole] count $n(\omega)$ between ARPES, optical and XAS experiments is striking, and is well
captured in the model calculations in Figs.~2(e)-(h). The HESW
plotted in Fig.~1(b) illustrates one important characteristic of
these curves to demonstrate the universality of the doping
dependence, but the detailed agreement is clearly much more
extensive.  This observation motivates our choice of the cut-off frequencies in
Fig.~2.  Since experimental and theoretical values are extracted in the same way, it is simplest to chose a doping-independent $\omega_c$ for each spectroscopy.  The natural
choice is the minimum spectral weight regions evident in Fig.~2, separating low and high energy scales.  These correspond to the waterfall region in
single particle spectra of ARPES and XAS or the isosbetic point in
optical spectra which is also the manifestation of the waterfall
effect as discussed in Ref.~\cite{tanmoysw}.  Our $\omega_c$ values are chosen
as average values which fall near this minimum.

\section{Intermediate coupling model of ASWT}
The theoretical calculations in Figs. 1 and 2 are based on
the QP-GW model\cite{tanmoysw,tanmoyZ}, an extension of our earlier
Hartree-Fock (HF) model of AFM gap collapse\cite{tanmoy07,tanmoy08,kusko,tanmoy06} to the
intermediate coupling regime by introducing a GW-like self-energy
correction.\cite{foottb,ft_sophi1,ft_sophi3,ft_sophi4}

The self-energy in QP-GW model is dominated by a broad peak in $\Sigma^{\prime\prime}$ which produces the `waterfall' effect\cite{markiewater,susmita} in the electronic dispersion by redistributing spectral weight into the coherent in-gap states and an incoherent residue of the undressed UHB and LHB. With underdoping, the in-gap states develop
a pseudogap which we model as a ($\pi,\pi$)-ordered spin density wave.
The doping evolution of both electron and hole-doped
cuprates is dominated by a magnetic gap collapse near optimal
doping.\cite{tanmoy07,tanmoy08}  The present calculations are obtained with the
same parameter sets as in Ref.~\onlinecite{tanmoysw}; in particular the doping
dependence of U is shown in Fig.~5 of that publication.
Our analysis identifies two main factors that cause ASWT. Firstly, the
pseudogap collapses with doping, shifting the optical MIR peak to low
energies while transferring weight to the Drude peak.  Secondly, the
residual incoherent weight associated with the Hubbard bands decreases
with doping\cite{tanmoysw} due to decrease in magnon scattering.
This is reflected in the doping dependence of the peak in $\Sigma^{\prime\prime}$.
The strength of this peak can be measured by the area under the $\Sigma^{\prime\prime}$ curve, Fig.~3.  This gives a direct measure of the tendency of the spectrum to split into
coherent and incoherent parts, and hence a measure of the weight of
the Hubbard bands.  Fig.~3 shows this quantity as a function of doping
above and below the Fermi level for both NCCO and LSCO. In both materials,
$\int\Sigma^{\prime\prime}~d\omega$ below $E_F$, seen in ARPES, shows a
much faster fall-off with doping.  This fast fall-off seems to terminate
around $x\sim 0.20-0.25$ close to the point where the HESW extrapolates to
zero, $x_{UHB}=1/\beta\sim 0.25$. This is also close to the doping where
AFM order ends in a critical point, suggesting an intimate connection
between the decrease of magnon scattering and the collapse of the AFM gap.
The good agreement between experiment and theory suggest that ASWT is predominantly associated with electron-electron interaction.\cite{phon}

The unusual doping dependence of the experimental $W_{UHB}$ in Fig. 1b can be understood within our model as follows.  The magnetic gap collapses near $x\sim$0.2 for both electron\cite{kusko,tanmoy06} and hole doped case\cite{tanmoy08}, and beyond this doping there is at most only a weak dip in the density of states, indicating a separation of the band into two components -- now coherent and incoherent parts.  However, since we work with fixed cutoff, we count all empty states in the band above $\omega_c$ as part of the UHB.  These change slowly with doping, decreasing linearly to zero at $x=0.2$.  Hence the break in slope indicates the magnetic gap collapse.

\begin{figure}[top]
\rotatebox{270}{\scalebox{0.5}{\includegraphics{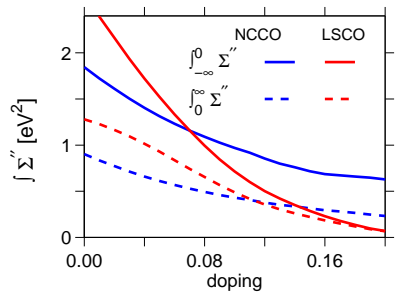}}}
\caption{(color online) Integrated imaginary part of
the calculated self-energy as a function of doping.} \label{sumrule}
\end{figure}

\begin{figure}[top]
\rotatebox{270}{\scalebox{0.4}{\includegraphics{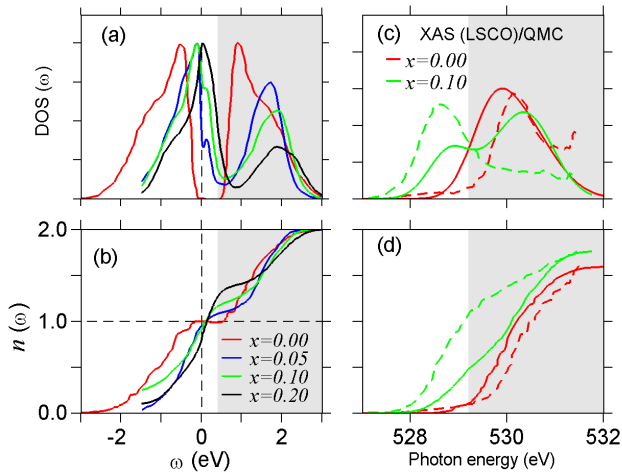}}}
\caption{(color online)
(a) DOS computed in QMC \cite{jarrell,maier}. All results are normalized
to their peak values. (b): Corresponding electron number $n(\omega)$, the
integral of the DOS, normalized to 2 for a full band.  The QMC results for
some of the dopings are not available at higher energies below $E_F$, so
we normalize $n(\omega)$ to $(1-x)$ at $E_F$. (c) and (d) QMC DOS with
experimental broadening and K-edge energy shift (solid lines) is compared
with experimental data (dashed lines)\cite{Pellegrin,peets} as in
Fig.~2(d-h).
} \label{sumrulef}
\end{figure}

\section{Discussion}
To better understand the failure of QMC calculations with fixed $U=8t$ to explain the observed ASWT, in
Figure 4(a-b) we plot the DOS and the associated electron count calculated in QMC\cite{jarrell,maier}.  For $\omega_c=0.4$~eV, close to the DOS minimum\cite{PhPh}, $W_{UHB}$ is in good agreement with the exact
diagonalization results for the corresponding $U=8t$,\cite{foot4}  blue stars in
Fig.~1(b), but has a considerably weaker falloff than found in experiment.  Note that the same result would follow by choosing a doping dependent $\omega_c$ pinned to the DOS minimum.
Consistent with this, we carried out a similar analysis of the XAS spectra
based on QMC-based DOS with doping-independent $U=8t$ in Figs.~4(c-d).
The QMC spectra (solid lines) are not consistent with the
experimental results, clearly overestimating the weight of the UHB for
finite dopings. Similar conclusions were reached in
Refs.~\onlinecite{peets,millis}.  Note that the mean-field result is similar: for $U=8t$, the gap collapse would be shifted to much higher doping $x\sim 0.43$.\cite{kusko}

Since the bare $U$ should be doping independent, the apparent doping dependence of the effective $U$ in our model arises to compensate for interactions not included in the underlying calculation. We have been able explain this doping dependence as due to long-range Coulomb screening\cite{tanmoysw}.  Alternatively, it should be noted that the doping dependence of $U$ can also be significantly reduced by going to a three-band model\cite{MBRIXS,DMFT2}. Indeed, it is common practice in the LDA+U literature to try to calculate a screened U by incorporating interactions involving other bands or longer range Coulomb interactions.  In this sense, our result is a natural extension to incorporate the doping dependence of this screening, which is particularly important near the metal-insulator transition.\cite{Anzai}

However, there is an ongoing debate on this issue that we would like to address.  Some screening is present within the one band Hubbard model, and it is important to see whether the full doping dependence of U could be understood on the basis of a more exact treatment of the Hubbard model -- i.e., whether the physics of cuprates can be fully understood within a single-band Hubbard model.  Clearly, as more correlations are added the doping dependence of the effective $U$ systematically decreases from Hartree-Fock calculations\cite{kusko} to the present QP-GW model, to recent DMFT calculations that can successfully describe the doping evolution of the cuprates with fixed-$U$ models\cite{comanac,millis,DMFT1}.  However, neither exact diagonalization nor QMC with fixed $U=8t$ capture the ASWT, Fig.~1(b), and the doping dependence of $U$ was not found in recent Gutzwiller calculations\cite{RM7}.  Indeed, by comparing the experimental results with exact diagonalization calculations, Fig.~1(b), a value $U<5t$ at finite $x$ is indicated, consistent with our results.  One way to reconcile the DMFT with the QMC and exact diagonalization results might be to note that both of the latter calculations are for a pure Hubbard model, neglecting band structure effects by restricting the overlap to nearest neighbor only.  Hence, it will be necessary to include at least a $t'$ in the exact diagonalization and QMC calculations to ensure that all three calculations converge on a common behavior for the one-band Hubbard model.

\section{Conclusions}
In conclusion, we have shown that the spectral weight of the UHB [LHB for
electron-doped cuprates] collapses with doping at a rate much faster than
can be explained in a $t-J$ or $U=\infty$ Hubbard model.  Such a fast
falloff would seem to require a real Mott gap collapse consistent with an
intermediate coupling $U <W$ scenario. We find that the rate of ASWT is universal --
the same across several spectroscopies and many different cuprates.
The plot of HESW vs doping in Fig.~1(b) provides a unique signature of the effective Hubbard $U$ in these materials.

\begin{acknowledgments}
This work is supported by the Division of Materials Science and
Engineering, US Department of Energy, under contract DE-FG02-07ER46352,
and benefited from the allocation of supercomputer time at NERSC,
Northeastern University's Advanced Scientific Computation Center
(ASCC).  RSM acknowledges support of a Marie Curie
Grant PIIF-GA-2008-220790 SOQCS.
\end{acknowledgments}


\begin{thebibliography}{99}
\bibitem{foot3}Similar effects are found for charge transfer insulators.\cite{EMS}
\bibitem{foottb} Doping independent tight-binding parameters have
been employed within a rigid band picture. A more realistic treatment of
doping effects on electronic states (see,
e.g., Refs.~\onlinecite{ft_sophi1,ft_sophi3,ft_sophi4}) was not
undertaken. However, we expect the rigid band model to be a good
approximation for doping away from the Cu-O planes.
\bibitem{foot4} This is also consistent with the QP-GW model: for a fixed $U=8t$ the pseudogap collapses
at an electron-doping $x\sim$0.43 in NCCO, compared to 0.19 for the doping dependent $U$.\cite{kusko}
\end{thebibliography}
\end{document}